\documentclass[conference]{IEEEtran}
\IEEEoverridecommandlockouts
\usepackage{cite}
\usepackage{amsmath,amssymb,amsfonts}
\usepackage{algorithmic}
\usepackage{graphicx}
\usepackage{textcomp}
\usepackage{xcolor}
\usepackage{tabularx} % for 'tabularx' environment

\usepackage{mdwlist}
\usepackage[ruled,vlined,shortend, linesnumbered]{algorithm2e} % algorithm
\usepackage[]{algorithmic} % algorithm
\usepackage{booktabs}
\usepackage{color}

\usepackage{amsthm,amssymb}
\usepackage{amsmath}
\usepackage[font=small]{caption}
\usepackage[skip=3pt]{subcaption}
\usepackage{multirow}
\usepackage{mathrsfs}
\usepackage{amsbsy}
\usepackage[mathscr]{eucal} %% For \mathscr
\usepackage[flushleft]{threeparttable}
\usepackage[hidelinks]{hyperref}
\usepackage{comment}
\usepackage{paralist}
\usepackage{pifont}% http://ctan.org/pkg/pifont

\usepackage{bbding} % for the \CheckmarkBold

\def\BibTeX{{\rm B\kern-.05em{\sc i\kern-.025em b}\kern-.08em
    T\kern-.1667em\lower.7ex\hbox{E}\kern-.125emX}}

\newcommand{\hide}[1]{}

\newcommand{\bit}{\begin{compactitem}}
\newcommand{\eit}{\end{compactitem}}
\newcommand{\ben}{\begin{compactenum}}
\newcommand{\een}{\end{compactenum}}

\begin{document}

\title{A WLAV-based Robust Hybrid State Estimation using Circuit-theoretic Approach
%\thanks{© 2020 IEEE. Personal use of this material is permitted. Permission from IEEE must be obtained for all other uses, in any current or future media, including reprinting/republishing this material for advertising or promotional purposes, creating new collective works, for resale or redistribution to servers or lists, or reuse of any copyrighted component of this work in other works.}
}

\author{Shimiao~Li,
	 Amritanshu~Pandey,
	 Larry~Pileggi\\
	Dept. of Electrical and Computer Engineering\\
	Carnegie Mellon University% <-this % stops a space
%\thanks{Authors are with the Department
%		of Electrical and Computer Engineering, Carnegie Mellon University, Pittsburgh,
%		PA, 15213}% <-this % stops a space
\thanks{This paper has been accepted by 2021 IEEE PES General Meeting. }
}

%\author{Shimiao Li, Bryan Hooi, Amritanshu Pandey, Christos Faloutsos, Larry Pileggi}
\maketitle

\begin{abstract}
	For reliable and secure power grid operation, AC state-estimation (ACSE) must provide certain guarantees of convergence while being resilient against bad-data. This paper develops a circuit-theoretic weighted least absolute value (WLAV) based hybrid ACSE that satisfies these needs to overcome some of the limitations of existing ACSE methods. Hybrid refers to the inclusion of RTU and PMU measurement data, and the use of the LAV objective function enables automatic rejection of bad data while providing clear identification of suspicious measurements from the sparse residual vector. Taking advantage of linear construction of the measurement models in circuit-theoretic approach, the proposed hybrid SE is formulated as a LP problem with guaranteed convergence. To address efficiency, we further develop problem-specific heuristics for fast convergence. To validate the efficacy of the proposed approach, we run ACSE on large cases and compare the results against WLS-based algorithms. We further demonstrate the advantages of our solution methodology over standard commercial LP solvers through comparison of runtime and convergence performance.
\end{abstract}


\section{Introduction}
\label{sec:Introduction}
Today’s central grid operators run AC state estimation (ACSE) at minute scale to gain situational awareness of the grid states. Network topology and grid measurements are inputs and grid states that include complex bus voltages and angles are produced as output. Furthermore, the output serves as the input to other critical grid analyses such as optimal power flow (OPF) and contingency analysis. It follows that for secure and reliable grid operation and control, robust ACSE is critical.

To ensure robustness, ACSE needs some resilience against faulty input measurements to promote reliable estimation. The grid today is highly susceptible to erroneous data from conventional sensors and networked PMUs due to disturbances (measurement device error, communication error) and malicious attacks. As a result, many methods have been proposed to make ACSE resilient from bad data. Most common amongst these are hypothesis tests that are based on post-processing of the weighted least square (WLS) problem residuals \cite{traditional-BDI},\cite{robustSE-reweight},\cite{robustSE-modify}. Within these algorithms, the identification of bad data is followed by adjusting suspicious measurements and rerunning ACSE until residuals are satisfactory. Possible ways of adjustments include removing bad data, modifying \cite{robustSE-modify} bad data, or iterative reweighted least square method \cite{robustSE-reweight}, where weights on suspicious measurements are set to lower values. Clearly, the iterative re-running of the ACSE problem incurs a computational burden. Other robust estimators include least median of squares \cite{robustSE-median}, yet it is not widely adopted due to numerical difficulty in handling the median.

Built upon the assumption that bad data are typically sparse, researchers have also proposed the use of weighted least absolute value (WLAV) in the objective function \cite{LAVSE-dc-soliman}\cite{LAVSE-dc-Vidya}\cite{LAVSE-PMU-abur}\cite{LAVSE-PMU-abur2}. The WLAV based methods enable automatic rejection of bad data, eliminating the need of iterative re-runs while providing clear identification of suspicious measurements from the sparse residual vector. Despite this desirable property, the non-differentiable L1-norm terms in the WLAV approach introduces additional complexity. At present, two approaches to WLAV-ACSE exist: i) those that solve highly non-convex ACSE due to inclusion of power measurements from conventional RTUs \cite{LAVSE-dc-Singh-splitform}; and ii) those that are formulated as a linear programming (LP) problem \cite{LAVSE-PMU-abur}\cite{LAVSE-PMU-abur2} with the assumption that network only consists of PMUs. These convex WLAV formulations based solely on PMUs take advantage of linear relationship between phasor data and grid states; however, assuming PMU-observability of the network is highly unrealistic for grids today.

To be readily applicable to grids today, ACSE must include conventional RTUs. Unfortunately, most methods that include conventional RTUs (hybrid or otherwise) correspond to ACSE methods with highly complex and non-convex solution space that presents convergence challenges and high residuals. To address these problems, there has been some work on convex relaxations of ACSE methods (applicable to WLAV as well). For instance, \cite{convexSE-SDP-weng} convexifies the problem by reformulating the voltage magnitude $|V|$, phase angle $\delta$, and product of $V$ and $I$, in semidefinite programming (SDP) form, through matrix transformation; however the method fails to scale effectively to large scale networks. More recently to overcome the convergence and high residual limitations, the researchers in \cite{SUGAR-SE-Li},\cite{SUGAR-SE-Alex} proposed a circuit theoretic ACSE approach that reduces the problem to an equality constrained quadratic programming (QP) problem by creating linear measurement models following a ‘sensitivity’ based mapping of conventional grid measurements. This approach provides a convex hybrid ACSE forumulation that is scalable to large networks, but it is not implicitly resilient to bad data.

In this paper we extend the work in \cite{SUGAR-SE-Li},\cite{SUGAR-SE-Alex} to create a convex weighted least absolute value (WLAV) based hybrid SE. Exploiting the linear modeling framework for both conventional RTUs and PMUs in \cite{SUGAR-SE-Li},\cite{SUGAR-SE-Alex}, we formulate the hybrid SE as a linear programming (LP) problem that provides implicit bad data detection. Also to solve the problem efficiently, this paper develops circuit-theoretic solution methodology with primal-dual interior point (PDIP) approach \cite{PDIP-L1norm} that is aided by heuristics for fast convergence on large-scale systems. To demonstrate the efficacy of this approach, results are presented for ACSE experiments on large networks to compare efficiency and performance against the simplex method and other general-purpose PDIP solvers.

\section{Circuit Formulation of SE Problem}
\label{sec:bkg}

\subsection{Equivalent Circuit Formulation (ECF)}
A circuit-theoretic formulation for power flow and grid optimizations was developed in \cite{SUGAR-pf}, \cite{SUGAR-opf}. Instead of describing components with power-based ‘PQV’ parameters, this framework models each component within the power grid as an equivalent circuit characterized by its I-V relationship. The I-V relationships can represent both transmission and distribution grids without loss of generality and, when applied to SE, can include measurement data naturally without incurring the non-linearity associated with state  variables. For computational analyticity, the complex relationships are split into real and imaginary sub-circuits whose nodes correspond to power system buses. %Table \ref{tab:comparison} shows a simple comparison between the traditional PQV formulation and ECF. 

% \begin{table}[htbp]
% %\centering
% \caption{Comparison between formulations}
% \label{tab:comparison}
% \begin{tabularx}{\linewidth}{lll}
% \toprule
%   {\bf Property} & {\bf PQV } &{\bf ECF} \\ 
% \midrule
% \begin{tabular}[c]{@{}l@{}} Coordinate \end{tabular} &  \begin{tabular}[c]{@{}l@{}} Polar \end{tabular}  &  \begin{tabular}[c]{@{}l@{}} Rectangular \end{tabular}\\
% \midrule
% \begin{tabular}[c]{@{}l@{}} State variables \end{tabular} &  \begin{tabular}[c]{@{}l@{}} $|V|,\theta$ \end{tabular}  &  \begin{tabular}[c]{@{}l@{}} Rectangular \end{tabular}\\
% \midrule
% \begin{tabular}[c]{@{}l@{}} Network balance \end{tabular} &  \begin{tabular}[c]{@{}l@{}} Zero power mismatch \end{tabular}  &  \begin{tabular}[c]{@{}l@{}} Zero current mismatch \end{tabular}\\
% \midrule
% \begin{tabular}[c]{@{}l@{}} Governing equations \end{tabular} &  \begin{tabular}[c]{@{}l@{}} Power balance at buses \end{tabular}  &  \begin{tabular}[c]{@{}l@{}} KCL* equations at buses \end{tabular}\\
% \bottomrule\\
% \end{tabularx}
% \footnotesize{*KCL refers to Kirchhoff's Current Law}
% \end{table}

\subsection{Incorporating PMU and RTU} \label{sec:models}
% \begin{figure}[h]
% 	\centering
% 	\includegraphics[width=1\linewidth]{substitution}
% 	\caption[]{Substitution theorem transforms the power system to a circuit system with measurement models.}
% 	\label{fig:substitution}
% \end{figure}
Today, critical states of bulk energy systems (BES) are sensed by conventional remote terminal units (RTUs) and networked phasor measurement units (PMUs) and these measured parameters are continuously sent to grid control rooms through secure telemetry. We have previously shown in \cite{SUGAR-pf} that the complete power grid model can be represented as an aggregated equivalent circuit. Now to include RTU and PMU measurements into this circuit model, we apply Substitution Theorem that posits that any measured element in the grid can be replaced with voltage and current sensor measurements from that component or the flow element without changing the overall solution. However, due to inherent noise in the measurement devices, for each measurement model we include an additional slack injection variable, to satisfy Kirchhoff’s current law (KCL).
We have previously shown in \cite{SUGAR-SE-Li},\cite{SUGAR-SE-Alex} that with linear measurements models, the ECF framework maps the entire measurement dependent power grid model into a linear equivalent circuit  representation of the entire system.

\begin{figure}[h]
     \centering
     \begin{subfigure}[h]{\linewidth}
         \centering         \includegraphics[width=0.6\linewidth]{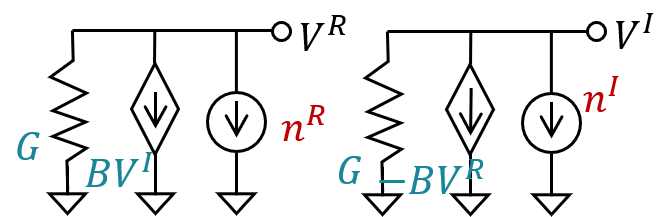}
         \caption{Linear RTU model: measurements are mapped to sensitivities.}
         \label{fig:RTU}
     \end{subfigure}
     \hfill
     \begin{subfigure}[h]{\linewidth}
         \centering         \includegraphics[width=0.5\linewidth]{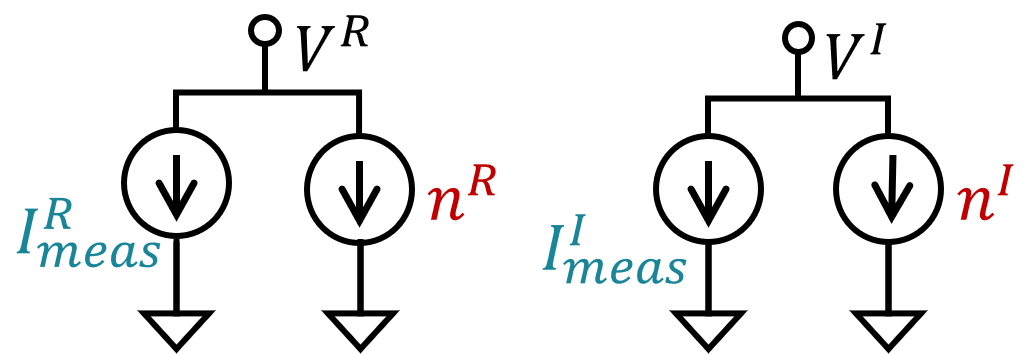}
         \caption{Linear PMU model: measurements have a linear nature under ECF.}
         \label{fig:PMU}
     \end{subfigure}
        \caption{RTU and PMU injection models.}
        \label{fig:models}
\end{figure}

 We briefly describe the construction of circuit-theoretic measurement models for RTUs and PMUs here (see \cite{SUGAR-SE-Li},\cite{SUGAR-SE-Alex} for details). Figure \ref{fig:RTU} represents the measurement circuit model for an RTU. In the model, the original injection measurements (voltage magnitude $|V|$, and power injection $P,Q$) are mapped into sensitivities ($G=\frac{P}{|V|^2},B=\frac{Q}{|V|^2}$) such that the I-V relationship is satisfied to obtain the linear circuit model. PMU measurements ($I^{real},I^{imag},V^{real},V^{imag}$) are intrinsically linear under a rectangular coordinate framework and the linear circuit model is characterized by independent current and voltage sources (see Figure \ref{fig:PMU}). Models for line flow measurements from RTU and PMU have been created in a similar way. Meanwhile, to account for possible measurement errors, the models have included noise terms, or slack variables, which are to be minimized.

\section{WLAV-SE definition and solution}
\subsection{Problem Definition}
\label{sec:LAV SE}
Even though existing circuit-theoretic WLS-based ACSE methods \cite{SUGAR-SE-Li} provide robust convergence guarantees, they lack an implicit bad-data detection capability that is becoming increasingly critical due to higher likelihood of coordinated attacks on grid measurements and naturally occurring faulty meters. Today, WLS-based SE applies a post-processing based method to perform bad data analysis by hypothesis test. This requires examination of residuals followed by updating and rerunning of SE in case any suspicious measurements are determined. Under the presence of bad data, such a methodology even with the existence of closed-form solution can face high computational burden. Moreover, since they are prone to false negatives, these methods can have undetected bad measurements that adversely impact the SE solution. 

To enhance the intrinsic reliability of the SE solution and eliminate the need for rerunning SE recursively, this paper develops a circuit-theoretic weighted least-absolute value (WLAV) estimator. By enforcing sparse value for residuals, this method can implicitly detect bad-data without the need for recursive re-runs of SE. Moreover, with the sparse residual vector capturing the pattern and impact of (sparse) measurement error, the solution to WLAV based SE is less corrupted and more reliable. And while existing WLAV SE methods that include both PMUs and RTUs have a non-convex complex solution space, we formulate a \textit{convex}
WLAV SE method with the linear measurement models described in Section \ref{sec:models}. Mathematical formulation is described as follows:

\begin{subequations}
\begin{equation}
    \min_{x,n} ||Wn||_1
\end{equation}
  s.t. KCL equations:\\
  PMU bus $i$:
  \begin{equation}
    Y_i^R x+I_{meas,i}^R+n_i^R=0
    \label{con 1st}
  \end{equation}
    \begin{equation}
    Y_i^I x+I_{meas,i}^I+n_i^I=0
  \end{equation}
  RTU bus $j$:
  \begin{equation}
    Y_j^R x+G_j V_j^R+B_jV_j^I+n_j^R=0
  \end{equation}
    \begin{equation}
Y_j^I x+G_jV_j^I-B_jV_j^R+n_j^I=0
  \end{equation}
  Zero-injection bus $k$: 
  \begin{equation}
  Y_k^R x=0
  \label{equ:ZI real}
  \end{equation}
  \begin{equation}
  Y_k^I x=0
  \label{equ:ZI imag}
  \end{equation}
  PMU line flow on line $p$:
  \begin{equation}
      Y_{line,p}^R x+I_{meas,p}^R+n_{line,p}^R=0
  \end{equation}
    \begin{equation}
Y_{line,p}^I x+I_{meas,p}^I+n_{line,p}^I=0
  \end{equation}
  RTU line flow on line q:
  \begin{equation}
      Y_{line,q}^R x+G_{line,q}V_{end,q}^R+B_{line,q}V_{end,q}^I+n_{line,q}^R=0
  \end{equation}
  \begin{equation}
Y_{line,q}^I x+G_{line,q}V_{end,1}^R-B_{line,q} V_{line,q}^R+n_{line,q}^I=0
    \label{con end}
  \end{equation}
  \label{prob: WLAV def}
\end{subequations}

\noindent where $W$ is a diagonal matrix of weights, $x$ is the state vector containing real and imaginary voltage variables, $R/I$ refers to real/imaginary, $Y$ is the admittance matrix, and $V_{end}$ refers to the bus where the line flow measurement device is metering.

Clearly, this is mathematically a linear programming (LP) problem for hybrid state estimation. The convex nature of the problem dispels any concerns regarding convergence issues even with conventional measurements. And built on the assumption that bad data are sparse, the solution to this problem automatically rejects them to return a reliable estimation. We now evaluate various solution methodologies for this problem and propose an algorithm with superior speed and robustness. %show that the proposed algorithm in Section \ref{sec:LAV SE} is superior to other known approaches in terms of speed and robustness.

\subsection{Fast and Robust Solution Methodology}\label{sec:solver method}

This section describes our solution methodology for efficiently solving the problem described in (\ref{prob: WLAV def}). It applies a  primal-dual interior point (PDIP) algorithm \cite{PDIP-L1norm} with novel problem-specific heuristics to achieve robust and fast convergence. This approach has superior convergence properties over simplex-based \cite{simplex} algorithms for large-scale problems. The Simplex method is generally better suited for smaller problems since it traverses through a set of vertices of the feasible space until the optimal solution is found. However, as the problem size increases, the number of vertices grow exponentially, making the method impractical for larger networks.

%Existing methods for solving the problem above are varied \cite{LP-solvers}. Examples include simplex method , active set algorithm , interior-point (IP) method \cite{PDIP-L1norm}, etc. Among them the commonly used in existing LAV ACSE is simplex method which traverses a series of vertices of the feasible region until reaching the optimal solution. This method takes advantage of geometry of the mathematical problem. Yet the efficiency and numerical stability decreases for large scale problems. In the worst-case, the algorithm has to visit most, if not all, of the vertices for optimality. As the problem size increases, the number of vertices grows exponentially, causing high computational burden. 
%
In the PDIP approach, to obtain the optimal solution, we solve the set of perturbed KKT conditions that are necessary and sufficient for optimality under convexity and strong duality. Compared with the exponential computational complexity of the Simplex method, this approach has proven to be effective on large-scale problems \cite{PDIP-L1norm}, with its worst-case complexity being polynomial to problem dimension. 

To apply PDIP algorithm to Problem (\ref{prob: WLAV def}), a standard transformation \cite{LAVSE-dc-Vidya},\cite{PDIP-L1norm} is performed to convert the non-differentiable L1 terms into differentiable format as below:

\begin{subequations}
\begin{equation}
\min_{x,n,t}w^Tt
\end{equation}
$$ s.t. \quad (\ref{con 1st})-(\ref{con end})$$
\begin{equation}
    |n|\preceq t
    \label{con neq}
\end{equation}
\end{subequations}

%\noindent Alternatively, another transformation, as in \cite{LAVSE-dc-Singh-splitform}\cite{LAVSE-dc-Abur-splitform}\cite{LAVSE-PMU-abur}\cite{LAVSE-PMU-abur2}, achieves the same goal by splitting the non-differentiable L1-terms into 2 non-negative variables such that $n=n^+-n^-,x=x^+-x^-,||n||_1=n^++n^-$ and $[n^+,n^-,x^+,x^-]^T\succeq 0$. Yet, this format lacks model interoperability and not used in our paper. 
%

Standard toolboxes can be used to solve the problem described in (2). These include CVXOPT 
\cite{CVXOPT}, SciPy \cite{scipy}, etc. Yet the (speed) performance of these solvers is limited for large grid cases. Therefore, to further improve the efficiency, we solve the perturbed KKT conditions with problem-specific limiting heuristics. Taking into account that the problem is convex and only local nonlinearity exists in the complementary slackness component of the perturbed KKT conditions, we apply simple step-limiting only on dual variables $\mu$ (corresponding to inequalities) and $t$ to make each iteration update faster and more efficient. This approach moves away from standard filter line-search algorithms \cite{CVXOPT} used by other generic tools. The proposed algorithm is shown in Algorithm \ref{alg:LP solver}.

\begin{algorithm}
	\caption{Variable limiting for solving LAV SE}
	\label{alg:LP solver}
	\KwIn{previous solution $\mu_{old}$,
	new solution $\mu,t,n$, step limit $d$}
	\KwOut{new solution $\mu,t$ after limiting}
	For each element $\mu_j$ in $\mu$:
	$$\Delta\mu_j=\mu_j-\mu_{old,j}$$
	$$dir = sign(\Delta\mu_j)$$
	$$h = \begin{cases} 1-\mu_{old,j} & dir \ge 0 \\ \mu_{old,j} & dir < 0 \end{cases}$$
    $$\mu_j=dir*\min(d,h)$$
	
	For each element $t_j$: 
	$$t_j = \begin{cases} 2|n_j| & |n_j|>t_j \\ t_j & else \end{cases}$$
\end{algorithm}

\subsection{Comparison and Interpretation of WLAV-SE versus WLS-SE}

We compared the proposed circuit-theoretic WLAV approach against the circuit-theoretic and traditional WLS approaches. Mathematically, 
%by reducing the redundant variables $n$, 
the proposed WLAV problem defined in Section \ref{sec:LAV SE} is equivalent to the following form: 
\begin{subequations}
\begin{equation}
    \min_x ||z'-Yx||_1
\end{equation}
  $$s.t. \quad (\ref{equ:ZI real})-(\ref{equ:ZI imag})$$
\end{subequations}
In contrast, our previous work \cite{SUGAR-SE-Li} has formulated hybrid SE as a quadratic programming (QP) problem %, the L2 norm based minimization does not have the property of automatically rejecting bad data as in LAV SE. 
%with an equivalent form as below:
\begin{subequations}
\begin{equation}
    \min_x ||z'-Yx||_2^2
\end{equation}
  $$s.t. \quad (\ref{equ:ZI real})-(\ref{equ:ZI imag})$$
\end{subequations}
And the classical 'PQV' based WLS-SE\cite{WLS-SE} is a direct minimization of measurement error, where $z$ is the vector of original measurements and $f$ is the non-linear relationship between $x$ and $z$:
\begin{equation}
    \min_x ||W(z-f(x))||_2^2
\end{equation}

\begin{figure}[htp]
\centering
\subfloat[Proposed WLAV based circuit-theoretic SE: LP problem (global optimum denotes the global optimum to the unconstrained problem that does not capture zero-injection node physics). Implicitly detects bad data.]
{%
  \includegraphics[clip,width=0.9\linewidth]{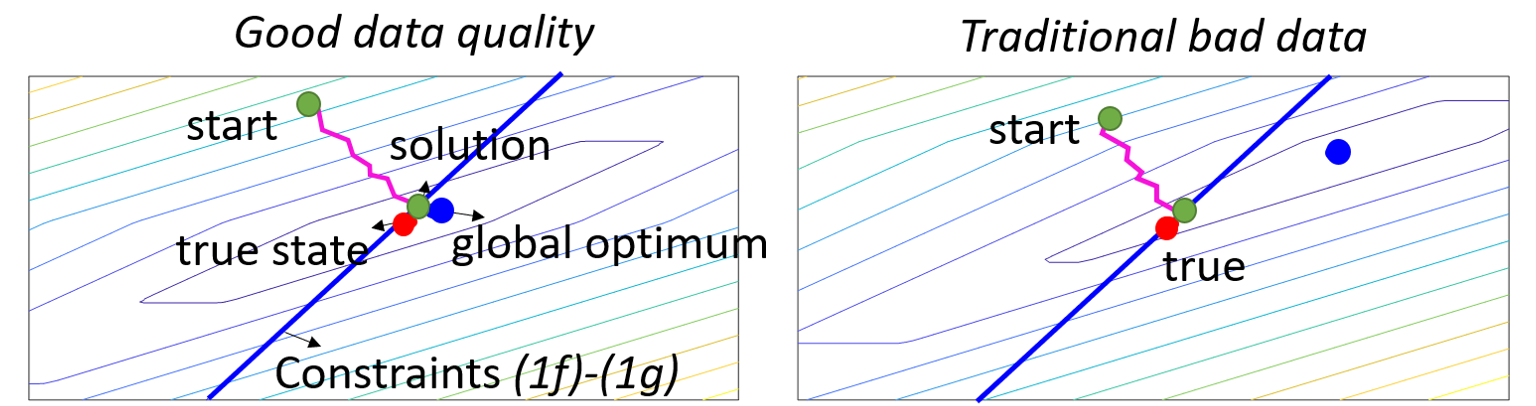}%
  \label{fig:sugar lav se}
}\\
\subfloat[L2-norm based circuit-theoretic SE \cite{SUGAR-SE-Li}: QP problem, has closed form solution. Does not detect bad-data.]{%
  \includegraphics[clip,width=0.9\linewidth]{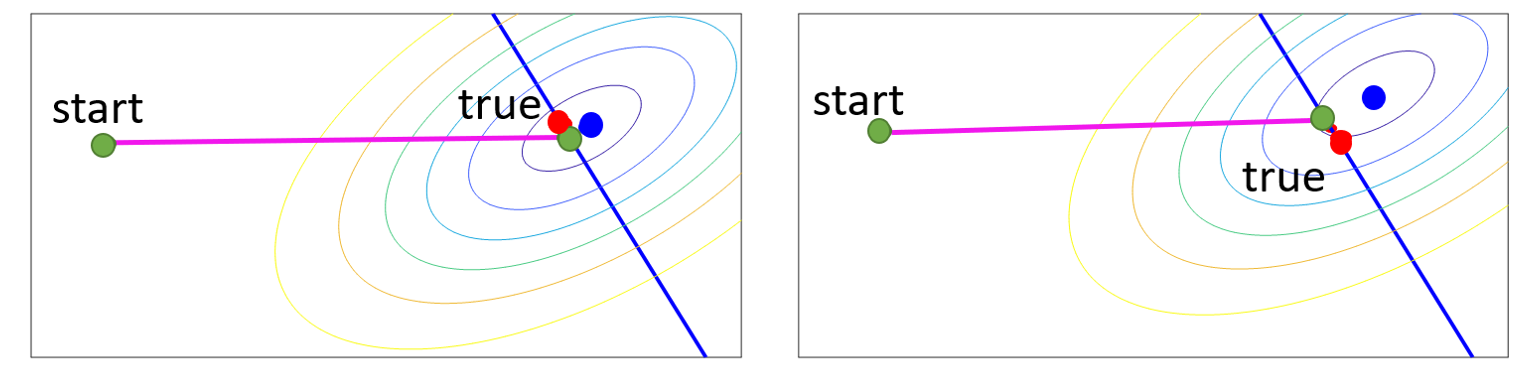}%
  \label{fig:sugar l2 se}
}\\
\subfloat[Classical WLS SE: non-convex problem, has convergence issues. Does not detect bad-data.]{%
  \includegraphics[clip,width=0.9\linewidth]{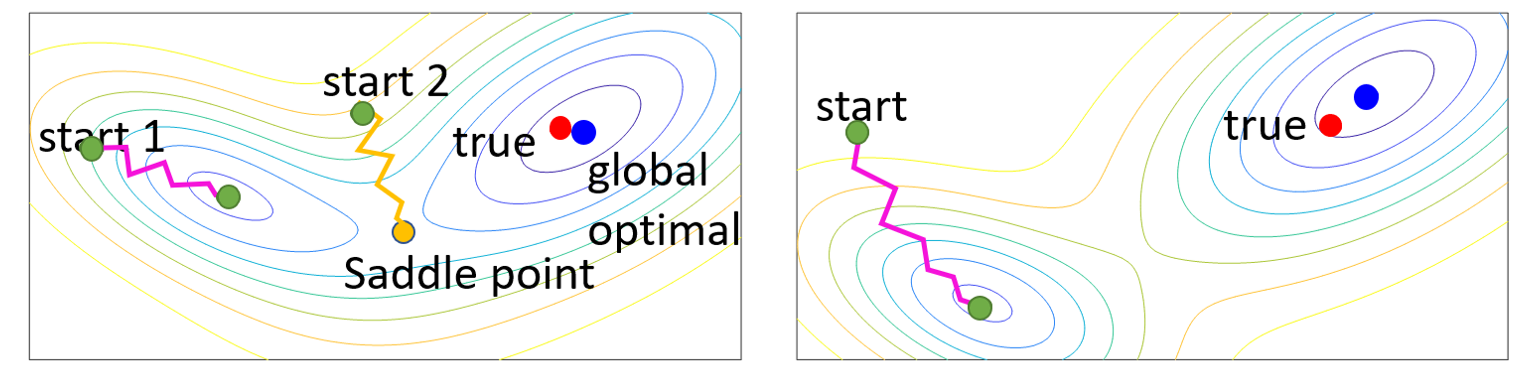}%
  \label{fig:wls se}
}
\caption{
Visual comparison of solution trajectory from an optimization viewpoint of a toy example.}
\label{fig:se_comparison}
\end{figure}

Figure \ref{fig:se_comparison} presents the solution trajectory of these different SE formulations.
Through comparison from an optimization viewpoint, it is clear that both our proposed method and our previous work \cite{SUGAR-SE-Li} benefit from convexity and the additional error-free zero-injection constraints. In contrast, the classical WLS method, which is non-convex and iterative, suffers from convergence to local minimum and saddle points. In addition to being convex, the proposed WLAV method provides additional benefits of implicit bad-data analysis, as justified by the results in later Section (see Table \ref{tab:comparison RMSE}). This not available with WLS-based methods in general.

\section{Results}
\label{sec:Results}
We conducted experiments to validate our proposed method for rejecting bad data. We constructed synthetic RTU injection and line flow measurements by adding Gaussian noise (std=0.001) to power flow solutions. To construct synthetic bad data, we added large deviations to a few randomly selected measurements. To evaluate the accuracy of the proposed estimation method, we use root mean square error $RMSE$ as the performance metric:
$$RMSE = \sqrt{||x_{est}-x_{true}||_2^2/n_{bus}}$$

Table \ref{table:case14} shows comparison results for IEEE 14 bus network, where bad measurements exist on bus 6 and 14. From general observation of nodal residuals that benefit from the LAV objective, our proposed method successfully identifies the location of bad data while the WLS method in \cite{SUGAR-SE-Li} fails to do so. Furthermore, smaller $RMSE$ shows the solution from WLAV SE is more accurate, validating its capability to reject bad data automatically.

\begin{table}[ht]
\centering
\caption{Magnitude of error term $|n_i|$ at each bus (CASE14)}
\begin{tabularx}{0.8\linewidth}{lll}
\toprule
{\bf Bus ID} & {\color{blue} \bf Proposed WLAV-SE} &{\bf WLS-SE\cite{SUGAR-SE-Li}}\\
\midrule
1 &  0 &  0\\
\midrule
2 &   0.009  &  0.124\\
\midrule
3 &  0  &  0.115\\
\midrule
4 &  0.065 &  0.121\\
\midrule
5 &  0  &  0.131\\
\midrule
6 (bad) &  \color{blue}1.095  &  0.313\\
\midrule
8 & 0.002   &  0.027\\
\midrule
9 & 0.0  &  0.090\\
\midrule
10 & 0.0  &  0.113\\
\midrule
11 & 0.0  &  0.059\\
\midrule
12 &  0.024  &  0.155\\
\midrule
13 &  0.0  &  0.199\\
\midrule
14 (bad) &  \color{blue}0.658  &  0.306\\
\midrule
{\bf RMSE} &  \color{blue}{\bf 0.042}  & {\bf 0.090} \\
\bottomrule
\end{tabularx}
\label{table:case14}
\end{table}

Next, we ran the proposed WLAV-SE on larger networks and compared our solution methodology against some standard optimization toolboxes \cite{CVXOPT}\cite{scipy}. In each test network, bad data was added on 5 randomly selected locations (data is available in \url{https://github.com/ohCindy/LAV-SE-synthetic-data.git}). 

\begin{table}[htbp]
%\centering
\caption{Comparison of $RMSE$}
\label{tab:comparison RMSE}
\begin{tabularx}{\linewidth}{lllll}
\toprule
CASE&\begin{tabular}[c]{@{}l@{}} WLS-SE\\(MATPOWER) \end{tabular} &  \begin{tabular}[c]{@{}l@{}} WLAV-SE\\ Simplex\\ (SciPy) \end{tabular}  &  \begin{tabular}[c]{@{}l@{}} WLAV-SE\\ IP\\ (CVXOPT)\end{tabular} &
\begin{tabular}[c]{@{}l@{}}\color{blue} WLAV-SE\\\color{blue}IP\\ \color{blue}(proposed) \end{tabular}\\
\midrule
14 & $9\times 10^{-2}$ & \begin{tabular}[c]{@{}l@{}} $4.2\times 10^{-2}$\\os:optimal \end{tabular} & \begin{tabular}[c]{@{}l@{}} $4.2\times 10^{-2}$\\os:optimal \end{tabular} & \color{blue}$4.2\times 10^{-2}$\\
\midrule
118 & $4.1\times 10^{-2}$ & \begin{tabular}[c]{@{}l@{}} $2.1\times 10^{-2}$\\os:optimal \end{tabular}  & \begin{tabular}[c]{@{}l@{}} $2.1\times 10^{-2}$\\os:optimal \end{tabular} &\color{blue}$2.1\times 10^{-2}$\\
\midrule
2383wp & $4.5\times 10^{-3}$ & \begin{tabular}[c]{@{}l@{}} $9.6\times 10^{-1}$\\os:infeasible \end{tabular}  & \begin{tabular}[c]{@{}l@{}} $5.9\times 10^{-4}$\\os:unknown \end{tabular} &\color{blue}$5.2\times 10^{-4}$\\
\midrule
6468rte & $4.7\times 10^{-3}$ & \begin{tabular}[c]{@{}l@{}} Fails \end{tabular}  & \begin{tabular}[c]{@{}l@{}} $6.7\times 10^{-4}$\\os:unknown \end{tabular} &\color{blue}$7.1\times 10^{-4}$\\
\midrule
9241pegase & $6.6\times 10^{-2}$ &Fails & \begin{tabular}[c]{@{}l@{}} $2\times 10^{-4}$\\os:unknown \end{tabular}  &\color{blue}$2.1\times 10^{-4}$\\
\bottomrule\\
\end{tabularx}
\footnotesize{
* We compare 4 methods: The traditional WLS based SE in MATPOWER, and WLAV-based SE defined in Section \ref{sec:LAV SE} implemented by 3 different LP solvers: Simplex method in SciPy toolbox, interior-point (IP) method in CVXOPT toolbox, and IP method by our proposed solver \\
* 'os' refers to optimization status reported by standard tooxbox. 'optimal' means optimal solution reached successfully, 'infeasible' means problem appears to be infeasible for the toolbox, 'unknown' means algorithm optimization terminated due to maximum iterations or numerical difficulties\\
}
\end{table}

Table \ref{tab:comparison RMSE} compares the accuracy of estimation, by using the $RMSE$ metric. The results show that WLAV-based ACSE (see columns 3,4,5 in Table \ref{tab:comparison RMSE}) provides superior estimates over WLS-based ACSE (see column 2 in Table \ref{tab:comparison RMSE}). %We further demonstrate superior performance over standard optimization toolboxes \cite{CVXOPT}\cite{scipy} (see columns 3,4 in Table \ref{tab:comparison RMSE}) that observe numerical difficulties or even divergence.

\begin{table}[htbp]
%\centering
\caption{Comparison of work-time(s) and iteration number}
\label{tab:comparison efficiency}
\begin{tabularx}{\linewidth}{llll}
\toprule
CASE& \begin{tabular}[c]{@{}l@{}} LAV, Simplex\\ (SciPy LP) \end{tabular}  &  \begin{tabular}[c]{@{}l@{}} LAV, IP\\ (CVXOPT)\end{tabular} &
\begin{tabular}[c]{@{}l@{}}\color{blue} LAV, IP\\ \color{blue}(proposed) \end{tabular}\\
\midrule
14& \begin{tabular}[c]{@{}l@{}} 0.125s, 129 iters \end{tabular}  &  \begin{tabular}[c]{@{}l@{}} 0.060s, 11 iters\end{tabular} &
\color{blue} \begin{tabular}[c]{@{}l@{}} 0.250s, 19 iters
\end{tabular}\\
\midrule
118& \begin{tabular}[c]{@{}l@{}} 8.897s, 3034 iters \end{tabular}  &  \begin{tabular}[c]{@{}l@{}} 1.110s, 16 iters\end{tabular} &
\color{blue}\begin{tabular}[c]{@{}l@{}} 1.026s, 47 iters
\end{tabular}\\
\midrule
2383wp& \begin{tabular}[c]{@{}l@{}} 23724s, 60237 iters \end{tabular}  &   \begin{tabular}[c]{@{}l@{}} 107s, 7 iters\end{tabular} &
\color{blue}\begin{tabular}[c]{@{}l@{}}51s, 37 iters
\end{tabular}\\
\midrule
6468rte& Fails  &  \begin{tabular}[c]{@{}l@{}} 750s, 9 iters\end{tabular} &
\color{blue}\begin{tabular}[c]{@{}l@{}} 262s, 34 iters
\end{tabular}\\
\midrule
9241pegase& Fails  &  \begin{tabular}[c]{@{}l@{}} 2538s, 9 iters\end{tabular} &
\color{blue}\begin{tabular}[c]{@{}l@{}}907s, 42 iters
\end{tabular}\\
\bottomrule\\
\end{tabularx}
\end{table}

Additional results in Table \ref{tab:comparison efficiency} 
%provide comparison of work time and iteration count. Results 
show that the proposed method has faster convergence for large cases when compared against baseline LP solvers. In general, CVXOPT takes fewer iterations to converge but each iteration is expensive due to the delicate computation of step-size and centering parameter to choose a proper search direction. In comparison, the proposed method, with simple heuristics, makes each iteration cheaper but requiring more iterations to converge. In contrast to both solvers, convergence of Simplex method significantly deteriorates when network size increases, which validates its limitation as stated in Section \ref{sec:solver method}.

\begin{figure}[h]
	\centering
	\includegraphics[width=0.77\linewidth]{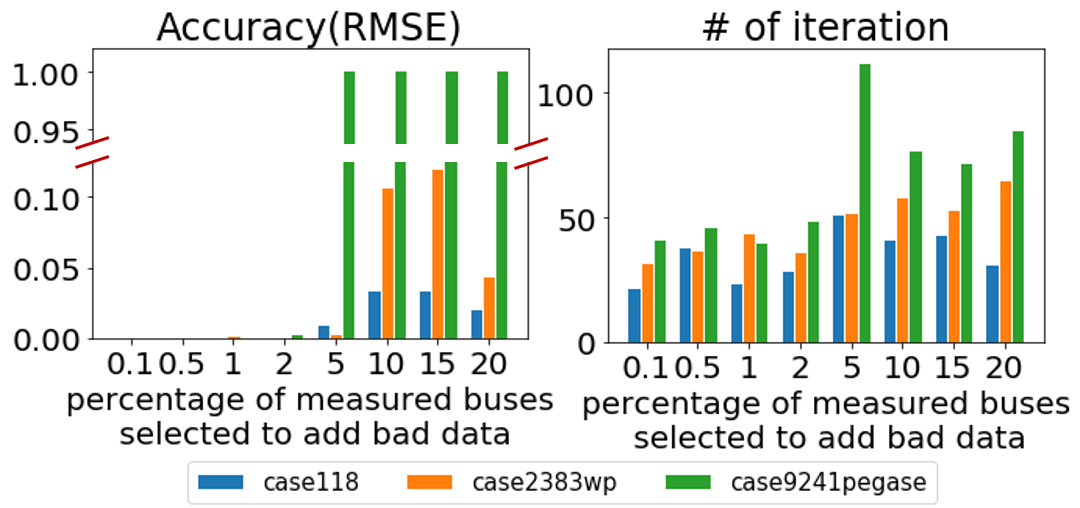}
	\caption[]{Accuracy and convergence efficiency under increasing population of bad data.}
	\label{fig:robustness}
\end{figure}
In the final experiment, we explore how our proposed method performs as the amount of bad data increases. Results (see Figure \ref{fig:robustness}) show that the convergence efficiency (\#iterations) remains stable under increasing bad-data samples, whereas the solution quality (accuracy) is stable for up to 5\% penetration of bad measurement buses and starts to degrade thereafter. With high penetration of bad data the property of sparse residual is not well preserved, which violates the basic assumption of all LAV methods. However, given that non-interactive bad data are highly sparse in reality, our approach is practical and can robustly reject sparse bad data in a manner that is superior to classical methods, like $r_N$-test\cite{traditional-BDI} which is less effective even for more than single bad data.

\section{Conclusion}
\label{sec:Conclusion}
This paper presents a hybrid state estimation built on a circuit-theoretic foundation and models. Benefits include: %The proposed ACSE method has the following benefits:
\begin{itemize}
    \item Guaranteed convergence due to formulation as a linear programming (LP) problem %with linear measurement models
    \item Intrinsic resilience against bad-data due to use of a least absolute value objective that produces sparse residual% and automatically rejects bad data
    \item Fast convergence due to the use of problem specific heuristics for the solution engine
\end{itemize}
%These properties provide significant advantages for situational awareness of future grids, which in addition to being larger in scale, are more robust against to anomalies in measurement data. 
Furthermore, the proposed SE approach that is directly applicable to both transmission and distribution grids is also generalizable to parallel compute techniques; a key need for tomorrow with merging transmission and distribution grids.

\section*{Acknowledgment}
This work was supported in part by the National Science Foundation (NSF) under contract ECCS-1800812. 

%The preferred spelling of the word ``acknowledgment'' in America is without 
%an ``e'' after the ``g''. Avoid the stilted expression ``one of us (R. B. 
%G.) thanks $\ldots$''. Instead, try ``R. B. G. thanks$\ldots$''. Put sponsor 
%acknowledgments in the unnumbered footnote on the first page.
\bibliographystyle{IEEEtran}
\bibliography{refbib}

\end{document}